# Creation of Tunable Homogeneous Thermal Cloak with Constant Conductivity

Tiancheng Han[1], Baowen Li[2], and Cheng-Wei Qiu[1]

[1]Department of Electrical and Computer Engineering, National University of Singapore, Kent Ridge 119620, Republic of Singapore.

[2]Department of Physics and Centre for Computational Science and Engineering, National University of Singapore, Singapore 117546, Republic of Singapore

**Abstract:** Invisible cloak has long captivated the popular conjecture and attracted intensive research in various communities of wave dynamics, e.g., optics, electromagnetics, acoustics, etc. However, their inhomogeneous and extreme parameters imposed by transformation-optic method will usually require challenging realization with metamaterials, resulting in narrow bandwidth, loss, polarization-dependence, etc. On the contrary, we demonstrate that tunable thermodynamic cloak can be achieved with homogeneous and finite conductivity only employing naturally available materials. The controlled localization of thermal distribution inside the coating layer has been presented, and it shows that an incomplete cloak can function perfectly. Practical realization of such homogeneous thermal cloak has been suggested by using two naturally occurring conductive materials, which provides an unprecedentedly plausible way to flexibly realize flexible thermal cloak and manipulate thermal flow.

Recently, many significant achievements of invisibility cloaking have been achieved hitherto, owing to pioneering theoretical proposals of Pendry[1] and Leohardt[2]. Since the ideal electromagnetic cloaks generally come with extremely complex constitutive parameters (inhomogeneous, anisotropic, and singular), many simplified strategies have been adopted, including reduced cloaks[3-6] and carpet cloaks[7-13]. To overcome the lateral shift problem of the carpet cloak designed by quasi-conformal mapping[14], ground-plane cloak constructed with calcite crystals[15,16] and one dimensional full-parameter cloak constructed with metamaterials[17] have been experimentally demonstrated recently. In addition to manipulation of electromagnetic wave[1-17], the theoretical tool of coordinate transformation has been extended to acoustic waves[18,19], matter waves[20,21], elastic waves[22], and heat flux[23-28].

On the basis of the invariance of heat conduction equation under coordinate transformations, transformation thermodynamics has provided a new method to manipulate heat flux at will[23]. The most attractive application is thermal cloaking: making the temperature of a certain region invariant. However, in an analog to electromagnetic cloaking[1,3], conventional thermal cloaking is dependent on its geometry, whose materials are usually inhomogeneous and singular[24-27], and in turn the practical applications of thermal cloak may be limited. More recently, an experiment has been reported to shield, concentrate, and invert heat current[28] by utilizing latex rubber and processed silicone.

This motivates us to establish the theoretical account and general design roadmap for creating realizable thermal cloak by using only homogeneous, non-singular, and natural conduction materials. The proposed novel thermal cloak is homogeneous, non-singular, independent on its geometrical size, and dominated by only anisotropy, which is distinguished from thermal cloaks[23-28] reported so far. More interesting is that, by judiciously selecting natural materials, a partially constructed cloak can perform perfectly. This unique functionality is enabled by the controllable thermal localization, e.g., most heat is confined in the vicinity of the cloak's outer boundary. Therefore an ultra-thin thermal cloak could be created, if a natural material with strong conduction anisotropy can be found or an effective material of highly anisotropic conduction can be constructed. Due to homogeneity and non-singularity, the proposed cloak may be fabricated by multilayer composition approach exploiting two naturally occurring materials throughout.

Thermal conduction is the movement of a heat flux from a high temperature region toward a low temperature region. For a steady state and no heat source, the thermal conduction equation can be written as $\nabla \cdot (\kappa \nabla T) = 0$, where $\kappa$ is the thermal conductivity, and $T$ is the temperature. Transformation thermodynamics has demonstrated that the conduction equation is invariant in its form under the coordinate transformation[23]. Specifically, in the transformed space, the thermal conduction equation can be written as $\nabla' \cdot (\vec{\kappa}' \nabla' T') = 0$. We can obtain

$$\vec{\kappa}' = \frac{\mathbf{A} \kappa \mathbf{A}^{\mathrm{T}}}{\det(\mathbf{A})}, \quad \text{with} \quad \mathbf{A} = \frac{\partial(x', y', z')}{\partial(x, y, z)} \tag{1}$$

Considering two dimensional case, where a circular region $(r \leq b)$ in original space

$(r,\theta,z)$ is compressed into an annular region $(a \leq r' \leq b)$ in physical space $(r',\theta',z')$. The transformation equation can be expressed as

$$r' = \frac{b-a}{b}r + a, \quad \theta' = \theta, \quad z' = z \tag{2}$$

Submitting Eq. (2) into Eq. (1) and assuming $\kappa = 1$, the conductivity of ideal cloak can be expressed as

$$\kappa'_r = \frac{r'-a}{r'}, \quad \kappa'_\theta = \frac{r'}{r'-a} \tag{3}$$

Clearly, the conductivity in Eq. (3) is spatially variable and has singularity at $r' = a$ ($\kappa'_r \to 0$ and $\kappa'_\theta \to \infty$), which is extremely difficult, if not impossible, to realize it. Apart from the ideal cloak, can we construct an advanced cloak with finite constant conductivity under the premise of maintaining perfect cloaking performance? Fortunately, this is feasible in thermodynamics. Examining Eq. (3), we can obtain the relationship $\kappa'_r \kappa'_\theta = 1$ and $0 \leq \kappa'_r = \frac{1}{\kappa'_\theta} \leq \frac{b-a}{b}$. By making $\kappa'_r = \frac{1}{\kappa'_\theta} = C$, where $C$ is a constant with $0 < C < \frac{b-a}{b}$, a homogeneous cloak without singularity is achieved. To validate the methodology, we introduce the rigorous theoretical analysis for such a homogeneous cloak, whose model is shown in Fig. 1. The conduction equation in cylindrical coordinate can be expanded as

$$\frac{\partial^2 T}{\partial r^2} + \frac{1}{r}\frac{\partial T}{\partial r} + \frac{l^2}{r^2}\frac{\partial^2 T}{\partial \theta^2} = 0 \tag{4}$$

where $l = 1$ for region I $(0 \leq r \leq a)$ and III $(r > b)$, $l = \sqrt{\kappa'_\theta/\kappa'_r}$ for region II $(a \leq r \leq b)$. Considering the symmetry relation $T(x,y) = T(x,-y)$, the temperature potential of three regions can be respectively expressed as

$$T_1 = \sum_{n=1}^{\infty} A_{2n-1} r^{2n-1} \cos(2n-1)\theta \tag{5a}$$

$$T_2 = \sum_{n=1}^{\infty} \left[ B_{2n-1} r^{(2n-1)l} + C_{2n-1} r^{-(2n-1)l} \right] \cos(2n-1)\theta \tag{5b}$$

$$T_3 = \sum_{n=1}^{\infty} \left[ D_{2n-1} r^{2n-1} + E_{2n-1} r^{-2n+1} \right] \cos(2n-1)\theta \tag{5c}$$

Owing to the temperature potential and the normal component of heat flux vector being continuous across the interfaces, we have

$$\begin{cases} T_1|_{r=a} = T_2|_{r=a}, & \nabla T_1|_{r=a} = \kappa'_r \nabla T_2|_{r=a} \\ T_2|_{r=b} = T_3|_{r=b}, & \kappa'_r \nabla T_2|_{r=b} = \nabla T_3|_{r=b} \end{cases} \quad (6)$$

Taking into account the boundary condition $T(x = \pm x_0) = \pm T_0$ (where $T_0$ is constant), we can obtain

$$T_1 = \frac{T_0}{x_0}\left(\frac{a}{b}\right)^{l-1} r\cos\theta, \quad T_2 = \frac{T_0}{x_0}\left(\frac{r}{b}\right)^{l-1} r\cos\theta, \quad T_3 = \frac{T_0}{x_0} r\cos\theta \quad (7)$$

Thus the temperature potential of three regions can be fully presented in Eq. (7). It can be seen that $l=1$ leads to $T_1 = T_2 = T_3$, corresponding to free space. For our homogeneous cloak, we need to control $l = \sqrt{\kappa'_\theta/\kappa'_r} = 1/C$, with $T_1/T_3 = (a/b)^{l-1}$ and $T_2/T_3 = (r/b)^{l-1}$. When $l$ is large enough, $T_1/T_3 \to 0$, implying that nearly no energy flows into the inner domain. It reveals that larger anisotropy corresponds to better performance, with the price of more difficult fabrication though. We want to have perfect performance ($T_1/T_3 \to 0$) and small anisotropy ($l \to 1$) simultaneously. However, it is not possible. In fact, each solution is some compromise between this two quantities, and each quantity can be improved by trading off the other one.

Full-wave simulations are carried out based on finite element method (FEM). Fig. 2 shows the temperature profiles for a thermal cloak with inner and outer radii of $a = 1\,\text{m}$ and $b = 2\,\text{m}$, respectively. Fig. 2(a) corresponds to the ideal cylindrical cloak described in Eq. (3), and Fig. 2(b)-(d) correspond to proposed homogeneous cloak with $C=0.1$, $C=0.2$, and $C=0.3$, respectively. As we can see, in Fig. 2(b)-(c), the heat fluxes travel around the inner domain and eventually returns to their original pathway. Therefore, the object inside the inner domain is protected from the invasion of external heat flux. Clearly, we have achieved advanced cloak with extremely simple parameters (finite constant conductivity), having performance as perfect as the ideal case.

When $C$ increases to 0.3, shown in Fig. 2(d), a small portion of thermal energy goes into the inner domain, which leads to an imperfect invisibility cloaking. Obviously, smaller $C$ (i.e. larger anisotropy) corresponds to better performance. To determine the maximal value of $C$, we may set

$T_1/T_3 = (a/b)^{l-1} \leq 0.1$, which means that the temperature potential in inner domain is negligible. Then we can obtain $\text{Max}(C) = \dfrac{\lg(a/b)}{\lg(a/b)-1}$. When $a = 1\,\text{m}$ and $b = 2\,\text{m}$, we can obtain $\text{Max}(C) = 0.23$. Clearly, as long as $C$ is smaller than Max($C$), nearly perfect performance can be achieved, as shown in Fig. 2(b) and (c).

To quantitatively examine cloaking performance with variance of anisotropy (denoted by $C$) and geometrical size (denoted by $b/a$), Fig. 3 (a) and (b) show the temperature gradient ($\nabla T$) of the inner region ($r<a$) as functions of $C$ and $b/a$. When the geometrical size is fixed ($a = 1\,\text{m}$ and $b = 2\,\text{m}$), temperature gradient as function of $C$ is demonstrated in Fig. 3(a). Clearly, nearly perfect performance can be achieved as $C$ is smaller than 0.23. When anisotropy is fixed ($\kappa'_r = 0.1$ and $\kappa'_\theta = 10$), i.e. $C=0.1$, temperature gradient as function of $b/a$ is demonstrated in Fig. 3(b). Obviously, good performance is kept until $b/a=1.3$, which means the cloaking shell is very thin. This is because $b/a=1.3$ corresponds to $\text{Max}(C) = 0.1$, and the fixed $C=0.1$ is not larger than Max($C$), thus good performance could still be achieved.

According to prediction of the theoretical equation $T_2/T_3 = (r/b)^{l-1}$, the temperature distribution in region II is more concentrated near outer boundary with the decrease of $C$, which has been shown from Fig. 2(b) to Fig. 2(d). To demonstrate this phenomenon clearly, Fig. 3(c) show the isothermal contour with different $C$. Obviously, the isothermal lines are more concentrated near outer boundary with the decrease of $C$. When $C=0.01$, nearly all of the energy in shell region is confined to the inner side of the outer boundary, which means an ultra-thin cloak with homogenous conductivity can be created.

Due to the proposed cloak with finite constant conductivity, it could be easily realized through alternating layered isotropic medium and only two types of isotropic materials (medium A and medium B) are needed throughout. The conductivities of medium A and medium B are defined as $\kappa_{A,B} = \kappa'_\theta \pm \sqrt{\kappa'^2_\theta - \kappa'_r \kappa'_\theta}$. Considering a thermal cloak with $a = 1\,\text{m}$, $b = 2\,\text{m}$, and $C=0.25$, i.e. $\kappa'_r = 0.25$ and $\kappa'_\theta = 4$, the temperature profile is shown in Fig. 4(a). The mesh formed by streamlines and isothermal values illustrates the deformation of the transformed space, which is curved smoothly around the central invisibility region. Fig. 4(b) presents the temperature profile for the multilayered cloak with homogeneous, nonsingular, and isotropic conductivities, which is believed to be the most advanced. The anisotropy in Fig. 4(a) has been removed by replacing the anisotropic material with two isotropic conductivities $\kappa_A = 7.87$ (thermal epoxy) and $\kappa_B = 0.13$ (natural latex rubber), as shown in Fig. 4(c). To validate the tunable property of the proposed cloak, we simulate the same size cloak with *C*=0.1, i.e. $\kappa'_r = 0.1$ and $\kappa'_\theta = 10$, as shown in Fig. 4(d). Since the energy mainly distributes near the outer boundary and is confined to the region ($0.15\,\text{m} \leq r \leq 0.2\,\text{m}$), we can construct a multilayered cloak in Fig. 4(e), where the multilayered material in no energy region ($0.1\,\text{m} \leq r \leq 0.15\,\text{m}$) is removed. The two constitutive materials of the thinner cloak are stainless steel ($\kappa_A = 20$) and wood ($\kappa_B = 0.05$), as shown in Fig. 4(f). Clearly, the multilayered cloak is as perfect as the ideal case. More importantly, one needs just two kinds of conductivities, in contrast to the singular and inhomogeneous cloak with 2*N* kinds of different conductivities[27].

In conclusion, we have proposed an advanced methodology for the design of thermal cloak with finite constant conductivity (without inhomogeneity and singularity), which drastically facilitates feasible realization and fabrication. The proposed cloak is independent on its geometrical size and is dominated by only anisotropy, which could be easily replaced by periodically alternating isotropic conductivities. Furthermore, given two isotropic conductivities, anisotropy can still be tunable in a large range by adjusting individual filling ratio, empowering many flexible recipes for using naturally occurring materials in thermal cloaking. In this connection, we demonstrate the possibility of creating ultra-thin thermal cloak by partial construction while maintaining perfect functionality. Theoretical analysis and full-wave simulations validate the advanced thermal cloak with utmost simple materials.

**References**

[1]  J. B. Pendry, D. Schurig, and D. R. Smith. Controlling electromagnetic fields. *Science* **312**, 1780–1782 (2006).

[2]  U. Leonhardt. Optical conformal mapping. *Science* **312**, 1777–1780 (2006).

[3] D. Schurig, J. J. Mock, B. J. Justice, et al. Metamaterial electromagnetic cloak at microwave frequencies. *Science* **314**, 977–980 (2006).

[4] B. Kanté, D. Germain, and A. Lustrac. Experimental demonstration of a nonmagnetic metamaterial cloak at microwave frequencies. *Phys. Rev. B* **80**, 201104 (2009).

[5] S. Xu, X. X. Chen, S. Xi, et al. Experimental demonstration of a free-space cylindrical cloak without superluminal propagation. *Phys. Rev. Lett.* **109**, 223903 (2012).

[6] H. Chen and B. Zheng. Broadband polygonal invisibility cloak for visible light. *Scientific Reports* **2**, 255 (2012).

[7] J. Li and J. B. Pendry. Hiding under the carpet: a new strategy for cloaking. *Phys. Rev. Lett.* **101**, 203901 (2008).

[8] R. Liu, C. Ji, J. J. Mock, J. Y. Chin, T. J. Cui, and D. R. Smith. Broadband ground-plane cloak. *Science* **323**, 366–369 (2009).

[9] H. F. Ma and T. J. Cui. Three-dimensional broadband ground-plane cloak made of metamaterials. *Nat. Commun.* **1**, 21 (2010).

[10] J. H. Lee, J. Blair, V. A. Tamma, Q. Wu, S. J. Rhee, C. J. Summers, and W. Park. Direct visualization of optical frequency invisibility cloak based on silicon nanorod array. *Optics Express* **17**, 12922 (2009).

[11] J. Valentine, J. Li, T. Zentgraf, G. Bartal, and X. Zhang. An optical cloak made of dielectrics. *Nat. Mater.* **8**, 568–571 (2009).

[12] L. H. Gabrielli, J. Cardenas, C. B. Poitras, and M. Lipson. Silicon nanostructure cloak operating at optical frequencies. *Nat. Photon.* **3**, 461–463 (2009).

[13] T. Ergin, N. Stenger, P. Brenner, J. B. Pendry, and M. Wegener. Three-dimensional invisibility cloak at optical wavelengths. *Science* **328**, 337–339 (2010).

[14] B. Zhang, T. Chan, and B.-I. Wu. Lateral shift makes a ground-plane cloak detectable. *Phys. Rev. Lett.* **104**, 233903 (2010).

[15] B. Zhang, Y. Luo, X. Liu, and G. Barbastathis. Macroscopic invisibility cloak for visible light. *Phys. Rev. Lett.* **106**, 033901 (2011).

[16] X. Chen, Y. Luo, J. Zhang, K. Jiang, J. B. Pendry, and S. Zhang. Macroscopic invisibility cloaking of visible light. *Nat. Commun.* **2**, 176 (2011).

[17] N. Landy and D. R. Smith. A full-parameter unidirectional metamaterial cloak for

microwaves. *Nat. Mater.* **12**, 25–28 (2013).

[18] H. Chen and C. T. Chan. Acoustic cloaking in three dimensions using acoustic metamaterials. *Appl. Phys. Lett.* **91**, 183518 (2007).

[19] S. Zhang, C. Xia, and N. Fang. Broadband acoustic cloak for ultrasound waves. *Phys. Rev. Lett.* **106**, 024301 (2011).

[20] A. Greenleaf, Y. Kurylev, M. Lassas, and G. Uhlmann. Isotropic transformation optics: approximate acoustic and quantum cloaking. *New J. Phys.* **10**, 115024 (2008).

[21] S. Zhang, D. A. Genov, C. Sun, and X. Zhang. Cloaking of matter waves. *Phys. Rev. Lett.* **100**, 123002 (2008).

[22] M. Brun, S. Guenneau, and A. B. Movchan. Achieving control of in-plane elastic waves. *Appl. Phys. Lett.* **94**, 061903 (2009).

[23] G. W. Milton, M. Briane, and J. R. Willis. On cloaking for elasticity and physical equations with a transformation invariant form. *New J. Phys.* **8**, 248 (2006).

[24] C. Fan, Y. Gao, and J. Huang. Shaped graded materials with an apparent negative thermal conductivity. *Appl. Phys. Lett.* **92**, 251907 (2008).

[25] T. Chen, C. N. Weng, and J. S. Chen. Cloak for curvilinearly anisotropic media in conduction. *Appl. Phys. Lett.* **93**, 114103 (2008).

[26] J. Li, Y. Gao, and J. Huang. A bifunctional cloak using transformation media. *J. Appl. Phys.* **108**, 074504 (2010).

[27] S. Guenneau, C. Amra, and D. Veynante. Transformation thermodynamics: cloaking and concentrating heat flux. *Opt. Express* **20**, 8207-8218 (2012).

[28] S. Narayana and Y. Sato. Heat flux manipulation with engineered thermal materials. *Phys. Rev. Lett.* **108**, 214303 (2012).

**Acknowledgements**

C.W.Q. acknowledges the Grant R-263-000-A23-232 administered by National University of Singapore. T.C.H. also acknowledges the support from the Southwest University (SWU112035).

**Figure captions**

**Figure 1** Schematic of the homogeneous thermal cloak.

**Figure 2** Temperature profile for a thermal cloak with $a=1\,\text{m}$ and $b=2\,\text{m}$. (a) Ideal conductivity described in Eq. (3). (b) $\kappa'_r = 0.1$ and $\kappa'_\theta = 10$. (c) $\kappa'_r = 0.2$ and $\kappa'_\theta = 5$. (d) $\kappa'_r = 0.3$ and $\kappa'_\theta = 3.3$. Isothermal lines are also represented with green color in panel.

**Figure 3** (a) Temperature gradient of the inner region ($r \leq a$) as function of $C$ with $a=1\,\text{m}$ and $b=2\,\text{m}$. (b) Temperature gradient of the inner region ($r \leq a$) as function of $b/a$ with $\kappa'_r = 0.1$ and $\kappa'_\theta = 10$. (c) Isothermal contour with different $C$ values at $a=1\,\text{m}$ and $b=2\,\text{m}$.

**Figure 4** Temperature profile for the thermal cloak with $a=1\,\text{m}$ and $b=2\,\text{m}$. (a) $C$=0.25. (b) The multilayered composition realization for the cloak in (a). (c) Close-up view of the multilayered cloak in (b) showing the constitutive materials available in nature. (d) $C$=0.1. (e) The multilayered composition realization for the cloak in (d). (f) Close-up view of the multilayered cloak in (e) showing the constitutive materials. Streamlines of thermal flux and isothermal are also represented with yellow and green colors in panel, respectively.

**Figures**

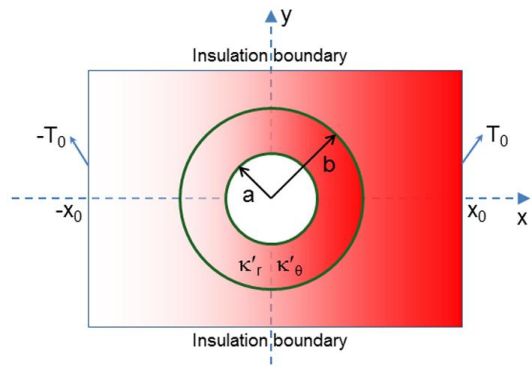

Figure 1

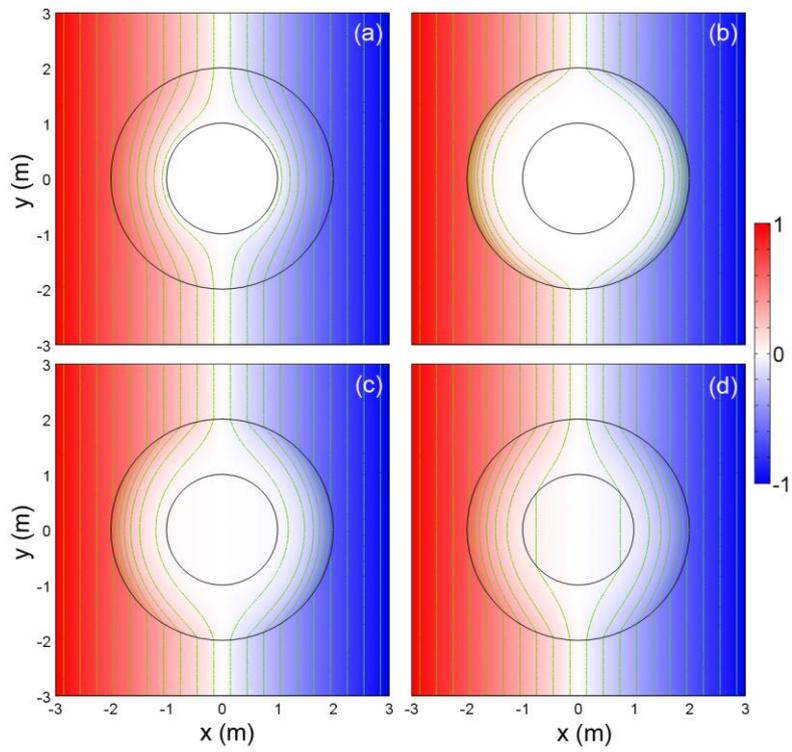

Figure 2

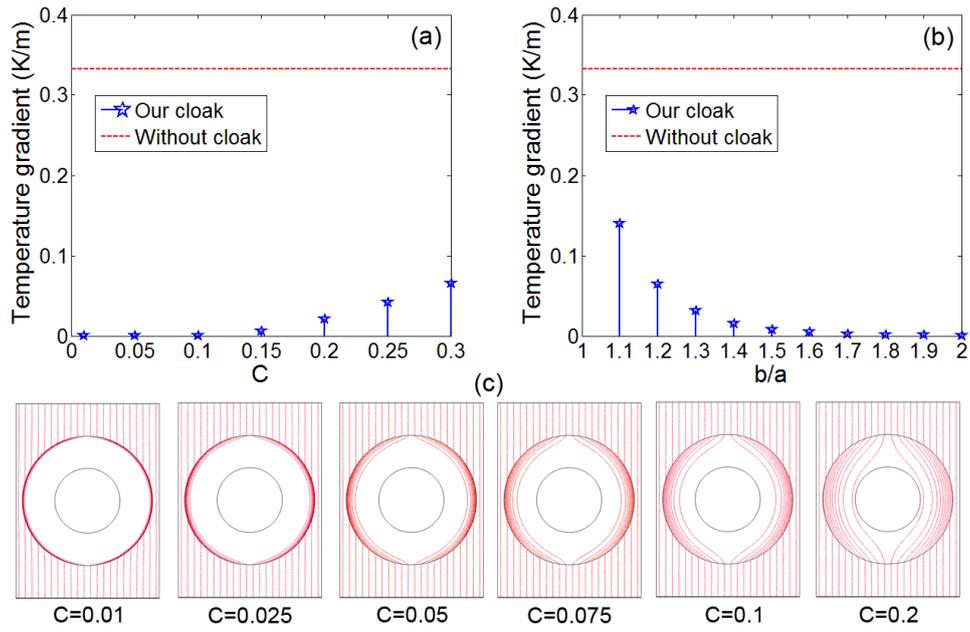

Figure 3

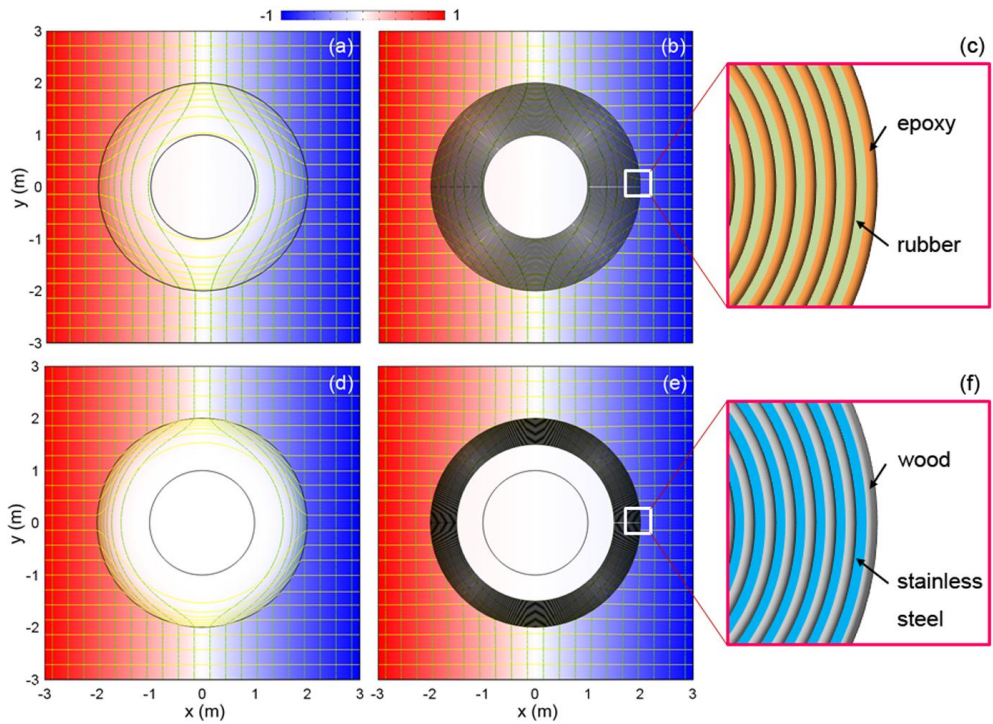

Figure 4